\documentclass{aa}
\usepackage{graphicx}
\def\etal{et al.\ }
\def\msol{$M_{\odot}$}
\def\bfig{\begin{figure}}
\def\efig{\end{figure}}
\def\degr{\hbox{$^\circ$}}
\def\arcmin{\hbox{$^\prime$}}
\def\arcsec{\hbox{$^{\prime\prime}$}}
\def\hr{\hbox{$^h$}}
\def\min{\hbox{$^m$}}
\def\sec{\hbox{$^s$}}

\begin{document}
\title{Multiband imaging observations of a prominent dust lane galaxy NGC 4370}   

\subtitle{}

   \author{M. K. Patil
        \inst{1}
        \and
S. K. Pandey
        \inst{2}
        \and
Ajit Kembhavi
        \inst{3}
	\and
D. K. Sahu
        \inst{4}
        }

   \institute{School   of   Physical  Sciences, S.R.T.M. University, Nanded, 431 606,  India\\
\email{patil@iucaa.ernet.in}
\and
School of  Studies in  Physics, Pt. Ravishankar Shukla University, Raipur,  492 010, India\\
\email{skp@iucaa.ernet.in}
\and
Inter University Centre for Astronomy \& Astrophysics (IUCAA), Post  Bag   4,  Ganeshkhind,  Pune,   411  007,  India\\
\email{akk@iucaa.ernet.in}
\and
Indian Institute of  Astrophysics, Bangalore, 560 034,  India\\
\email{dks@crest.ernet.in}
    }

  \date{Received / Accepted }

\abstract{
In this paper we present extinction properties of interstellar dust in a prominent dust lane galaxy NGC 4370 based on the optical broad band (BVRI) imaging observations taken from the Himalaya Chandra Telescope (HCT), Hanle and the near-IR (J,H,K$_s$) images taken from the 2MASS archive. NGC 4370 belongs to the Virgo cluster (VCC 0758) and form a non-interactive pair with NGC 4365 at 10$\arcmin$. NGC 4370 hosts a prominent dust lane running parallel to its optical major axis and is extended almost up to 1\arcmin. The extinction curve derived for NGC 4370 is found to run parallel to Galactic extinction curve, implying that the properties of dust in NGC 4370 are identical to those of the canonical grains in the Milky Way. The $R_V$ value is found to be equal to 2.85$\pm$0.05 and is consitent with the values reported for the dust lane galaxies. The total dust content of NGC 4370 estimated using optical extinction and IRAS flux densities are found to be equal to $4.4\times 10^4$ \msol and $2.0\times 10^5$ \msol, respectively. As regard to the origin of dust and ISM in this galaxy, the accumulated dust by this galaxy over its life-time is insufficient to account for the detected mass by optical means, which in turn imply that the ISM might have been acquired by the NGC 4370 through a merger like event. An attempt is also made to study the apparent spatial correspondence between the multiple phases of ISM, i.e., hot gas, warm gas and dust in this galaxy by obtaining  optical emission maps from narrow band imaging and diffuse X-ray emission map obtained from the analysis of \emph{Chandra} archival data. This analysis implies a physical connection between the dust and warm gas in terms of their physical co-existence and common origin too. 

}
\titlerunning{Multiband study of ISM in NGC 4370}
\authorrunning{M.K. Patil et al.}
   \maketitle

\section{Introduction}
Traditional views of early-type galaxies as simple structures, devoid of gas and dust has changed significantly in last few decades with the availability of advanced observing facilities across the electromagnetic spectrum. Now it is established beyond doubt that the early-type galaxies are not simple structures but host multi-phase interstellar medium and an ongoing star formation process (Goudfrooij 1999,  Trinchieri \& Goudfrooij 2002, Athey \& Bregman 2002, Patil \etal 2003, Humphrey \& Boute 2006 and the references there in). Further, it is believed that most of the elliptical galaxies we see today are the results of either merger interactions of two galaxies of comparable sizes, swallowing of a smaller galaxy by a larger companion or a grazing interaction from a nearby passing perturbing galaxy. Large attempts were made to understand the mechanism of formation history of early-type galaxies by studying their dynamics, kinematics and environmental effects. 

The dust-lane early-type galaxies have received a systematic attention after realising their importance in determining the three dimensional structure of galaxies and also for understanding the subsequent evolution of interstellar medium and the host galaxies (Bertola \& Galleta 1978). With the wealth of data available from very sensitive and powerful detectors, it has become clear that the dust features in galaxies act as tracer of the structure, stellar population dynamics, and post merger histories of the early-type galaxies to which they belong. Studying the properties of dust particles in the interstellar space can help us to understand the nature of grains themselves, the processes which govern their evolution and the way how dust grains respond to the light from the stars in their vicinity. The last property is due to the interaction of dust grains with the electromagnetic radiation, commonly called as an extinction, and depends strongly on the chemical composition and size distribution of the dust grains. To account for extinction, it is important to know the relative distribution of stars and gas within a galaxy. As individual stars cannot be resolved in galaxies at longer distance, therefore the foreground dust screen model has proved to be most successful model in examining the dust extinction properties in early-type galaxies (Brosch \& Loinger 1991, Goudfrooij \etal 1994b, Patil \etal 2007, Finkelman et al. 2008). In the present study we assume that the extinction of starlight in optical regime is caused by a turbulent screen in front of the emitting stars. 

Broadly speaking, interstellar extinction varies as a function of wavelength in such a way that the shorter the wavelength stronger is the extinction and vice-a-versa, and is generally represented by means of a graph called as {\it extinction curve}. The general shape of the extinction curve in the ultraviolet to near-IR (0.125 to 2.5 $\mu$m) region in our own Galaxy, the Milky Way, is generally characterised by a single parameter $R_V$ (Cardelli \etal 1988). The $R_V\,\left[=\frac{A_V}{E(B-V)}\right]$ is the measure of the total extinction in V band ($A_V$) to the selective extinction in B \& V bands ($E(B-V)=A_B-A_V$), and is assumed to correlate  with the average dust grain size. For the case of the Milky Way its typical value is equal to 3.1. Therefore, by studying nature of extinction curve in external galaxies it is possible to compare the dust grain sizes responsible for the extinction in optical pass band of the host galaxy.  

We have an ongoing programme of surface photometric analysis of a large sample of galaxies having dust lanes and or patches with a goal to establish a baseline of dust extinction properties in galaxies evolving in different environments and to compare those with that of the Milky Way. Here, we focus on the nearby early-type galaxy NGC 4370, classified as S0 in the Binggeli, Sandage \& Tammann (1985), de Vaucouleurs \& de Vaucouleurs (1964) and possess features more reminiscent of lenticular galaxies (Cinzano \& van der Marel 1994). NGC 4370 host a prominent dust lane running parallel to its optical major axis. This galaxy also hosts other phases of ISM like warm gas (Caldwell \etal 2003), hot gas (Bettoni \etal 2003) and substantial amount of neutral hydrogen ($H_I$) (Hutchmeir \& Richter 1986). Caldwell \etal (2003) on the basis of the higher order Balmer line-strength analysis demonstrate that age of NGC 4370 is about 3.7 Gyr, probably formed through a merger like event and have undergone a massive star formation. 

In this paper, we present the optical broad band (BVRI) and narrow band ($H_\alpha$) observations of NGC 4370 with an objective of studying dust extinction properties and physical association of dust with ``warm gas" and ``hot gas". In Section 2, we describe the optical observations and data reduction steps. The dust extinction properties, dust mass estimation is discussed in Section 3, Section 4 deals with the physical association of dust with other phases of ISM and the star formation process in NGC 4370. We summaries our results and discuss their implications briefly in Section 5.

\section{Observations and data analysis}
\begin{table*}[htb]
\caption[]{Global Parameters of NGC 4370}
\centering
\label{global}  
\scriptsize
\begin{minipage}[t]{0.5\linewidth}
{\begin{tabular*}{1.0\linewidth}%
{@{\extracolsep{\fill}}l l}\hline\hline
Parameter &	Value\\
\hline
RA (J2000.0) &	12\hr24\min55\sec\\
Dec (J2000.0) & +07\degr26\arcmin42\arcsec\\
Distance	& 11.20 Mpc\\
Morphological type: & S0\\
Size& 1.\arcmin50 $\times$ 0.\arcmin8\\
Magnitude & $M_B$ =-17.99; $B_T^0$=13.48; $L_B = 1.2\times10^9\, L_{\odot}$\\
V$_{helio}$ & 782 $\pm$ 7 km/s\\
Other names: &	VCC 0758; UGC 07492; PGC 040439\\
FIR flux densities (mJy) & {\emph f(12$\mu$m)= $110\pm 34$}; {\emph f(25$\mu$m)= $160 \pm 64$;} \\
& {\emph f(60$\mu$m)= $960 \pm 39$}; {\emph f(100$\mu$m)= $3260 \pm 87$}\\
HI (21 cm line)& 3.17 Jy\,km$^{-1}$ s$^{-1}$\\
\hline
\end{tabular*}}
\end{minipage}
\end{table*}

Deep CCD images of NGC 4370 were acquired during April 2005 using 2.0\,m Himalayan Chandra Telescope of Indian Astronomical Observatory (IAO), Hanle, India, in the Bessels B, V, R \& I broad band filters as well as the narrow band filter centred on $H_{\alpha}$ emission. The detailed global parameters of NGC 4370 are given in Table 1.  These observations were made using 1k$\times$1k Photometrics CCD camera with a total field of view $5\arcmin \times 5 \arcmin$, providing a pixel scale of 0.\arcsec285 pixel$^{-1}$. The exposure times were selected by taking care that the CCD pixels covering bright stars in the neighbourhood of galaxy should not get saturated. In order to achieve good signal-to-noise ratio, multiple frames of NGC 4370 were taken in each of the broad band filters B, V, R and I as well as the narrow band filter centred on $H_{\alpha}$ emission. Apart from science frames, several calibration frames like bias, twilight sky frames for correcting the non uniformity of response of CCD, were also taken in this observing run. For photometric calibration of the raw data, Landolt (1992) standard fields were observed.

The data was processed with IRAF\footnote{IRAF is a Image Reduction and Analysis Facility and is distributed by the National Optical Astronomy Observations (NOAO, which is operated by the Association of Universities, Inc. (AURU) under co-operative agreement with the National Science Foundation.}, following standard preprocessing steps like, bias subtraction, division by normalised twilight flat frame in each pass band, etc. Multiple frames taken in each pass band were geometrically aligned to an accuracy up to one tenth of a pixel by measuring centroids of several common stars in the galaxy frames and were then combined with median scaling to improve the S/N ratio. This median combination was also useful for  removing cosmic ray events in the object frames. The final co-added images were then flux-calibrated by applying the air-mass correction and instrumental offsets determined by analysing repeated observations of Landolt (1992) standard stars. The sky background in respective frame was estimated following the {\it box method} (Sahu \etal 1998) and was then subtracted from the science frame. 

\section{Isophotal Shape analysis}
\subsection{Surface photometric analysis}
The task {\it ELLIPSE} available within STSDAS\footnote{STSDAS is distributed by the Space Telescope Science Institute, operated by AURA, Inc., under NASA contract NAS 5-26555.} was used to quantify isophotal parameters of NGC 4370 in all the pass bands. {\it ELLIPSE} adopts the methodology described by Jedrzejewski (1987); i.e., for each semi major axis length $a$, the intensity $I(\phi)$ is azimuthally sampled along an elliptical path described by the initial guess parameters like isophote's centre (X,Y), ellipticity ($\epsilon$), and position angle ($\theta$). The intensity along the ellipse $I(\phi)$ is then expanded into a Fourier series as: 
\[
I(\phi) = I_0 +\sum_n \left[A_n\, sin(n\phi)\, +\, B_n\, cos(n\phi)\right]
\]

where $I_0$ is the intensity averaged over the ellipse, and $A_n$ and $B_n$ are the higher order Fourier coefficients. The best-fitting set of parameters $X, Y, \epsilon, \theta$ are those that minimises the sum of the squares of the residuals between the data and the fitted ellipse when the expansion is truncated to the first two coefficients (i.e. $A_n\, \&\, B_n >$ 3 will be zero). Non zero values of higher order coefficients (with $n \ge 3$) are due to the deviations of isophotes from being perfect ellipses. In practice, the third and fourth order coefficients are derived from the above equation by fixing the first and second order coefficients to their best-fit values. Here the third order coefficients ($A_3$ \& $B_3$) represent isophotes with three-fold deviations from ellipses (e.g., egg-shaped or heart-shaped), while the fourth order coefficient represents the four-fold deviations (e.g., rhomboidal or diamond shaped). For galaxies which are not distorted by interactions, the isophotes are ``disk'' shaped and are represented by positive values of $B_4$ ($B_4 > 0$), while the negative values of $B_4$ ($B_4 <0$) represents the``box'' shaped isophotes (Jedrzejewski 1987).

\begin{figure}[!htb]
\begin{centering}
\includegraphics[width=3.5in]{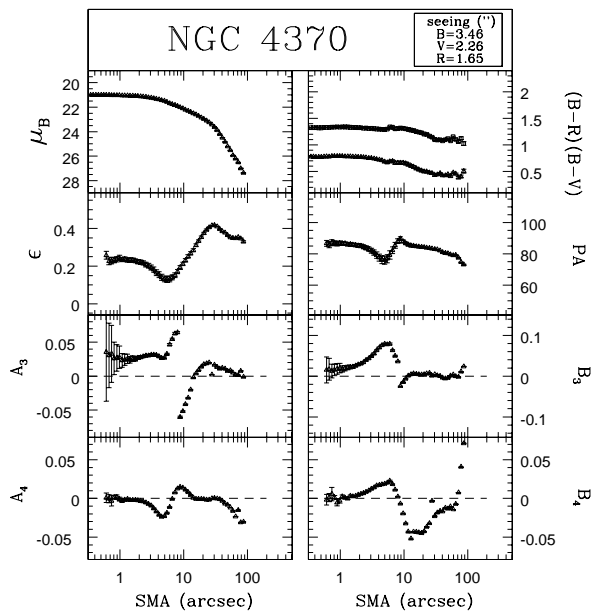}
\caption{Surface brightness profile and profiles of the structural parameters plotted as a function of radial distance.}
\label{srfphoto}
\end{centering}
\end{figure}

To avoid distortions in isophotal fitting due to the presence of foreground stars, other interfering objects and defects, etc., those regions were masked and neglected during the fitting process. Isophotal fitting was carried out by incrementing the semi major axis logarithmically, i.e. each isophote was calculated at a semi major axis 10\% longer than that of its preceding isophote. The output table generated during ellipse fitting was used to derive various  profiles such as, surface brightness profile, position angle, ellipticity, and higher order Fourier coefficients, all plotted as a function of radial distance from the centre of the galaxy. Figure \ref{srfphoto} shows the results derived from the isophotal shape analysis on NGC 4370 and are in good agreement with those reported by Caon \etal (1994). Large variations are seen in the profiles of higher order coefficients and are perhaps due to the presence of prominent dust lane in the target galaxy (Goudfrooij \etal 1994a). The colour profile was derived by carrying out ellipse fitting on different pass band images and by comparing the light distribution in the two different pass bands. 

\begin{figure}[!htb]
\begin{centering}
\includegraphics[width=2.75in]{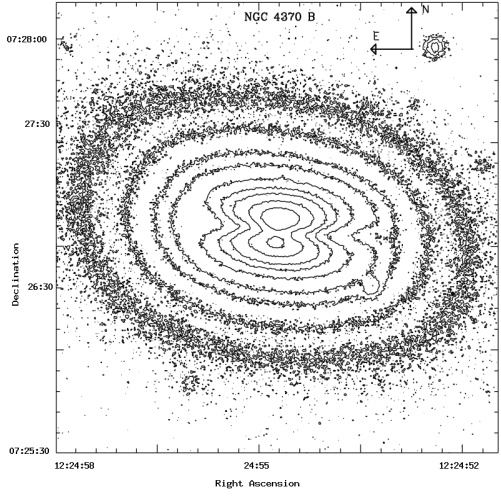}
\caption{Isointensity contours in the blue filter CCD image of NGC 4370.}
\label{contour}
\end{centering}
\end{figure}

\begin{figure}[!htb]
\begin{centering}
\includegraphics[width=2.75in]{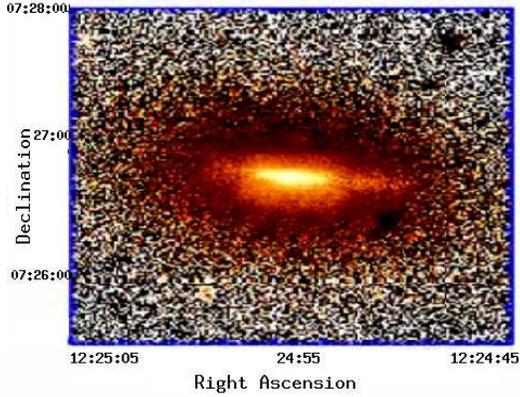}
\caption{(B-V) color index map delineating the prominent dust lane running along the optical major axis of NGC 4370}
\label{BVcmap}
\end{centering}
\end{figure}

\subsection{Dust extinction properties}
Though presence of prominent dust lane in NGC 4370 was reported by Bertola \etal (1988), van den Bergh \& Pierce (1990), Caon \etal (1994) and was also noticed from the isointensity contour map of NGC 4370 (Figure \ref{contour}), to determine exact orientation and extent of the dust lane in NGC 4370 we constructed colour index maps (B-V, B-R, B-Ks, etc.) using the geometrically aligned and seeing matched images in different pass bands. Here near IR  images (J,H,Ks) were taken from the archive of 2MASS observatory and were scaled to match with the images in optical pass bands for their different plate-scales. One of such colour index map (B-V) is shown in Figure \ref{BVcmap}, which delineats a prominent dust lane (bright shade) running along optical major axis of NGC 4370 and extending up to about one $arcmin$. Similar colour index maps were used to determine the morphology and extent of the dust distribution within NGC 4370 and were also used for calculating dust mass. 

To quantify the wavelength dependent nature of dust extinction in NGC 4370, one is expected to compare the light distribution in the observed galaxy with that of its dust free model. As the target galaxy is of early-type, in which the light distribution is fairly smooth, therefore, its dust free model can be easily constructed by fitting ellipses to the isophotes of the observed images (Brosch \& Loinger 1991, Goudfrooij \etal 1994b). Owing to this, we have fitted ellipses to the isointensity contours on images of NGC 4370 taken in different pass bands as discussed above, but here the dust occupied regions were masked and flagged off while fitting ellipses. Due to the strong dust obscuration in NGC 4370, ellipse fitting on optical images was not an easy task. We, therefore, first carried out the ellipse fitting on the longer pass band image i.e., $K_S$ pass band taken from the 2MASS archive, and the same centre coordinates were given while carrying out ellipse fitting on the images in optical pass bands. Before this near-IR images in J, H and $K_S$ pass bands were geometrically aligned and were scaled to match with the images in optical pass bands for their different plate-scales. A model image constructed using the best fitted ellipses was then subtracted from the original galaxy image and its residual image was generated. After examining the hidden features (dust occupied regions) in the residual image, those regions were flagged and rejected in the later run of the ellipse fitting.  

\begin{figure}[!htb]
\begin{centering}
\includegraphics[width=2.75in]{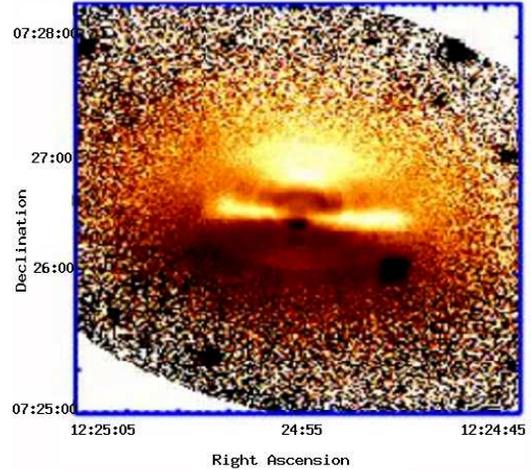}
\caption{Extinction map of NGC 4370 in B pass band, brighter shades represent the dust obscuration}
\label{Bext}
\end{centering}
\end{figure}

Dust free models generated above were then used to quantify the dust extinction $A_\lambda$ at wavelength $\lambda$ using the equation:
\[
A_\lambda = -2.5 \times log \left[\frac{I_{\lambda, original}}{I_{\lambda,model}}\right]
\]
where $I_{\lambda,original}$ is the observed intensity at a given point in the dust lane and $I_{\lambda,model}$ is the assumed input intensity of the starlight with no dust obscuration. It is to be noted that absolute flux calibration is not needed for this purpose. The extinction map of NGC 4370 in the B band is shown in Figure \ref{Bext} which represents a prominent dust lane running along the optical major axis of NGC 4370. The extinction in B band, excluding seeing affected central region, was found to vary from 0.88$\pm$0.03 to 0.36$\pm$0.03. The extinction maps generated in different pass bands were used to study the dust properties in NGC 4370. 

\begin{figure}[!htb]
\begin{centering}
\includegraphics[width=2.5in]{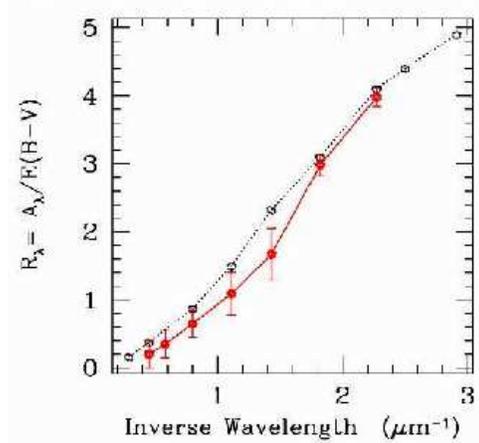}
\caption{Extinction curve derived for NGC 4370 (solid line) and for the Milky Way (dotted line) as a function of $\lambda^{-1}$}
\label{extncrv}
\end{centering}
\end{figure}

We then quantified total extinction $A_\lambda$ ($\lambda = B, V, R, I, $ etc.) in each pass band and used these values for deriving the extinction curve. Here, numerical values of $A_\lambda$ were extracted by sliding a 5$\times$5 box on the dust occupied region in each pass band and were used to estimate the local selective extinction $\left[ E(\lambda - V) = A_\lambda - A_V \right]$ as a function of position across the dust-occupied region. Then a regression fit was performed between local values of total extinction $A_\lambda$ to the selective extinction $E(B-V)$ (for details see Goudfrooij \etal 1994b, Sahu \etal 1998) and from the slopes of best fitted lines between ($A_\lambda$) and E(B-V), $R_\lambda$ values were obtained. These $R_\lambda$ values were used to derive the {\it extinction curve}, the wavelength dependence of the extinction, over optical pass-band B through the near-IR $K_s$. The extinction curve derived for NGC 4370 (Figure\,\ref{extncrv}) is found to run parallel to that for the Milky Way implying that the dust grains responsible for the extinction of stellar light in NGC 4370 have similar physical characteristics as that of the canonical grains in the Milky Way. The $R_V$ value derived in the case of NGC 4370 is 2.85$\pm$ 0.05 and is in good agreement with the value reported by Finkelman \etal (2008). The $R_V$ value in NGC 4370 is smaller than the average value of 3.1 in the Milky Way and imply that the dust grains responsible for the extinction of stellar light in the target galaxy are relatively smaller in size compared to those in the Milky Way.

Assuming that the chemical composition of the extragalactic dust grains is uniform throughout the galaxy and is similar to that in our Galaxy, it is possible for us to compute the relative dust grain size and the total dust content in the programme galaxy. Several models have been proposed in this regard, however, the two-component model (Mathis, Rumpl \& Nordsieck 1977) consisting of spherical silicate and graphite grains with an adequate mixture of sizes is appropriate to explain the observed extinction curves in the galaxies in our local neighbour hood including the Milky Way. This model assumes uncoated refractory grains having a power-law  size distribution of 
\[ n(a)da = n_H\, A_i\, a^{-3.5}\,da\]
where $a$ represents the grain size, $n_H$ is the hydrogen number density and $A_i$ is the abundance of component $i$ among the two. We used the upper and lower limits of grain size distribution at $a_+ = 0.25\,\mu$m and $a_-=0.005\,\mu$m, respectively. 

Then the relative size of the dust grains responsible for the extinction of star light in the optical region was estimated. This was done by comparing the wavelength-dependence of the extinction efficiency of spherical grains with that of the observed $R_\lambda$ values, which varies linearly with $\lambda^{-1}$. Thus, for a given value of dust extinction, the mis-match between the two extinction curves plotted for the target galaxy and the Milky Way is mainly because of the difference in relative grain size of the two galaxies. Thus, by shifting the observed extinction curve along $\lambda^{-1}$ axis till it best match with the Galactic extinction curve, we estimate the relative grain size. In the case of NGC 4370, the relative dust grain size is found to be about $\frac{<a>}{a_{Gal}} = 0.90 \pm 0.08$ and matches well with the value quoted by Finkelman \etal (2008).

\subsection{Dust mass estimation}

In order to estimate the total dust mass in target galaxy NGC 4370, we use the dust column density given by (Goudfrooij \etal 1994b)

\[
\Sigma_d = l_d \times \int_{a_-}^{a_+} \frac{4}{3} \pi a^3 \rho_d\, n(a)\, da
\]

where $\rho_d$ and $l_d$ represent the specific grain density and the dust column length along the line of sight, respectively. These values can be estimated from the measured total extinction and the extinction efficiency in the V band (for details see Goudfrooij \etal 1994b, Patil \etal 2007). The total dust mass was computed by integrating dust column density over the dust occupied region and using relation $M_d\,=\,\Sigma_d \times Area$, expressed in solar units. Using optical extinction measurement, we estimate the dust mass in NGC 4370 equal to $4.4\times 10^4$ M$_\odot$ and is in perfect agreement with the value reported by Finkelman \etal (2008). 

The dust mass was also estimated independently using the far-infrared emission from this galaxy as detected by the Infra-Red Astronomical Satellite (IRAS). For this we used the relation (Young \etal 1989)
\[
M_{dust} = 4.78\, S_{100} \, D^2\,\left[exp(143.88/T_{dust})-1\right]\, M_\odot
\]
where $S_{100}$ is the IRAS flux density at 100 $\mu$m in Jy taken from the catalogue of Knapp \etal (1989), D is distance in Mpc (assuming $H_0=70 \,km \,s^{-1} \,Mpc^{-1}$), and T$_{dust}$ is the dust temperature in K derived from the flux densities measured at 60 and 100 $\mu$m using $T_{dust} =(\frac{S_{60}}{S_{100}})^{0.4}$ (Young \etal 1989). We estimate the dust content in this galaxy using the IRAS flux densities to be $2.0\times 10^5$ M$_\odot$, roughly an order of magnitude more than that detected by optical method. This discrepancy in the two estimates is mainly due to the fact that optical extinction takes care of only that component of dust which have strongest effect on the starlight, whereas the IRAS flux density takes care of the emission originating even from the diffusely distributed component of dust. Since IRAS was insensitive to the cold dust with T\,$\le$\,20 K, emitting predominantly at wavelengths beyond 100\,$\mu$m, therefore this discrepancy may get enhanced even further when we incorporate flux densities at longer wavelengths, like that measured using ISO (Tsai \& Mathews 1996, Merluzzi 1998, Temi \etal 2004). The measurements carried out at sub-millimter wavelength using the SCUBA instrument have showed that nearly 1000 times more dust is likely to exist in the host galaxies in the cool form (Thomas et al. 2004; Stevens et al. 2005). Perhaps, the sub-millimeter observations detected by SCUBA orginates from the cool component of dust ($\sim$ 10-15 K) which was missed by IRAS as well as ISO.

\section{Other phases of ISM in NGC 4370}
\subsection{{\it ``Warm"} gas}
With a view to examine the association of dust with ionised gas, narrow band images of NGC 4370 centred around $H_\alpha+[NII]$ were taken using HCT with a total integration time of 3000\,s. The narrow band images were also processed in the same manner as explained above before generating the emission-line intensity images. The images in $H_\alpha+[NII]$ pass band were aligned with those in R pass band to an accuracy of few hundredths of a pixel. After co-adding these well aligned images in respective filters, the image with best seeing was convolved to match seeing PSF of the other band image (see Macchetto \etal 1996 for details). 

\begin{figure}[!htb]
\begin{centering}
\includegraphics[width=3.00in]{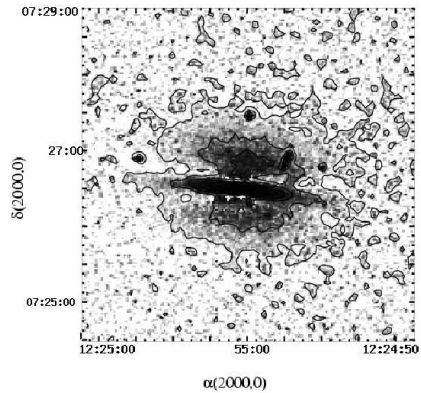}
\caption{Morphology of the ionised gas derived from narrow band observation, darker shade show emission region}
\label{Halpha}
\end{centering}
\end{figure}

Then intensity scale factor for R band image was determined by using the aligned and PSF matched images following the procedure outlined by Macchetto \etal (1996). For this, ellipses were fitted to the isophotes of the R band as well as the narrow-band images and a least square fit was carried out on the outer isophotes. This was essential to avoid the extra contribution coming from the ionised gas present in the dust occupied region. The slope of this fit was applied on the R band image as a scale factor and then the resultant image was subtracted from the narrow-band image to produce the pure ionised gas emission map. Figure\,~\ref{Halpha} shows morphology of the ionised (``warm") gas in NGC 4370 and delineate a close association of it with the dust. As the narrow band images were not acquired through the proper redshifted filters, therefore flux calibration of the $H_\alpha$ emission line was not possible.

\subsection{GALEX data}
\begin{figure}[!htb]
\includegraphics[width=1.7in]{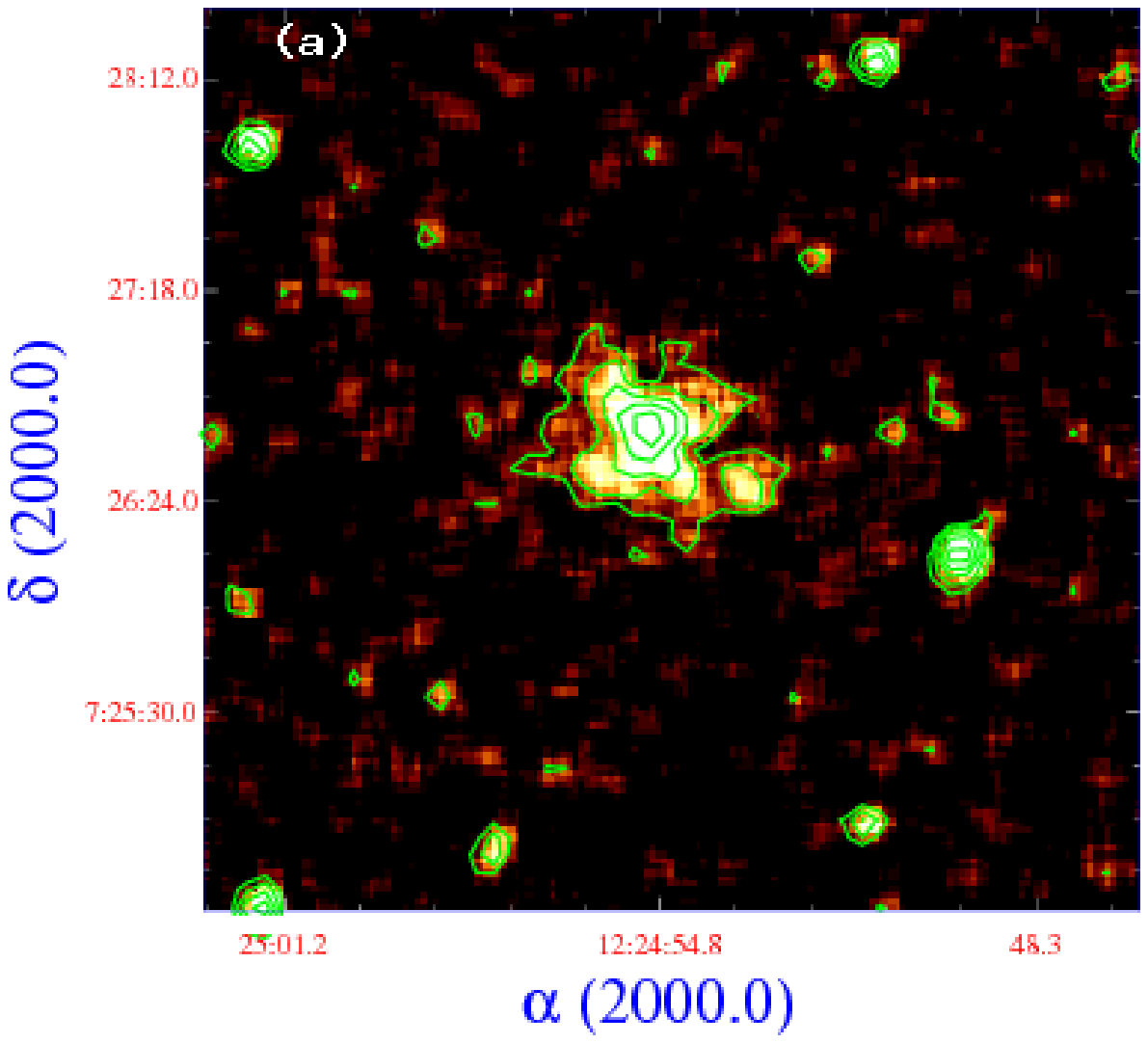}
\includegraphics[width=1.65in]{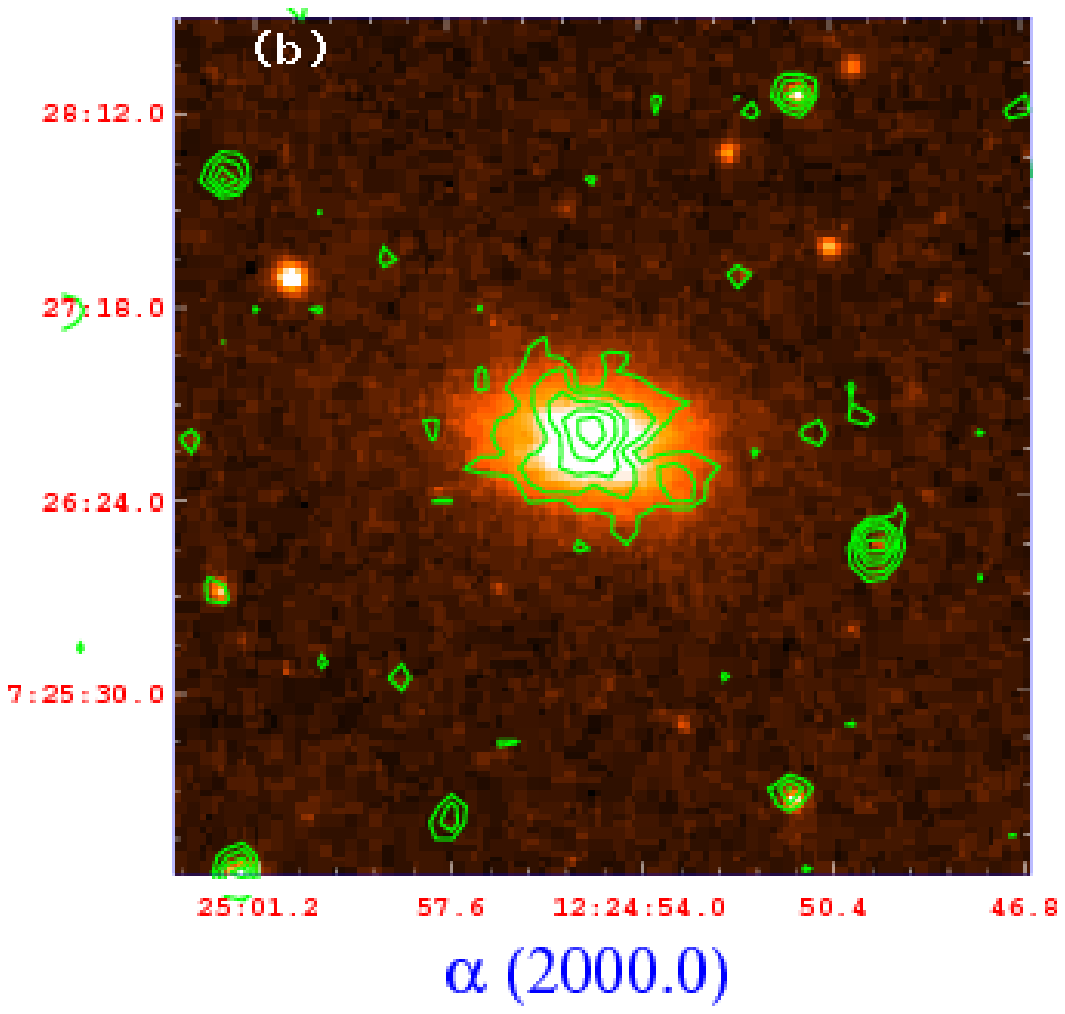}
\caption{(a) Near-UV map for NGC 4370 using GALEX data and overlaid are the contours (in log scale), (b) contours of Near-UV emission are overlaid on the DSS image}
\label{nuv}
\end{figure}

We have also used the {\it GALEX} images in the Near-UV and Far-UV pass bands with a view to trace the flux at UV wavelengths in NGC 4370 and to check its correspondence with the dust and other phases of ISM. Figure ~\ref{nuv}(a) shows the the Near-UV image of NGC 4370, superimposed over which are the isointensity contours. The same are overlaid on the DSS image, Figure\,~\ref{nuv}(b), which shows that the centre of the UV component matches with that of optical image. Here only near-UV image of GALEX have been used for tracing the UV flux in NGC 4370 due to its better signal-to-noise (S/N) ratio compared to the far-UV. The dust features are also discernible in the NUV image and are consistent with those seen in the optical counterpart. Here to enhance the appearance of the dust features in NUV image, we applied adaptive filter on it.
 
\subsection {{\it ``Hot"} gas}
X-ray emission from NGC 4370 was reported by Bettoni \etal (2003) using the data from Einstein and ROSAT space missions. However, because of their poor spatial resolutions it was not possible to compare its association with the optical counterpart. Owing to the superb spatial resolution of Chandra it is possible to examine association of dust with the hot (X-ray emitting) gas. In this regard we searched for the availability of X-ray data on NGC 4370 in the archive of {\it Chandra} space mission. We could not find X-ray data centred on NGC 4370, however, a 41\,ks X-ray image centred on NGC 4365 was available  in which NGC 4370 was sitting about 10$\arcmin$ away from the NGC 4365. The same image was used for examining morphology of hot gas in NGC 4370. This image was processed using CIAO 3.2.0 and calibration file CALDB 3.1.0. Event-2 file was first filtered to remove flaring period from this image using {\it lc\_clean.sl} script, which resulted in to a clean image of total integration time equal to 39.7\,ks. We then used CIAO {\em wavdetect} tool to generate the full energy range (0.3-10.0 keV) diffuse emission map of NGC 4370. Our analysis of X-ray data does not show obvious emission feature of hot gas in NGC 4370, perhaps due to the reason that NGC 4370 was lying near the edge of the ACIS-3 detector, where its response was not good. However, a deep X-ray image centred on the NGC 4370 may result in to the detection of hot gas in this galaxy. 

\subsection{Star Formation Rate}
Despite the fact that elliptical galaxies have processed most of their matter for star formation in the first few billion years from their formation, in some of the early-type galaxies evidence of current star formation activity have been reported (Caldwell 1984 and more recently by Sarzi \etal 2006, Yan \etal 2006). Star formation activity in this class of  galaxies depend on the availability and amount of cold gas in the galaxies. Recent surveys of ISM in early-type galaxies over the complete range of electromagnetic spectrum have amply demonstrated that most of the galaxies contain a large amount of ISM existing in all the known forms. NGC 4370 is also detected at 21cm emission line with a measured flux value of about 3.17 Jy\,km\,s$^{-1}$ giving rise to a total atomic gas content equal to 1.3$\times$ 10$^8$ \msol  available for the star formation. Thronson \& Bally (1987) by studying the IRAS colour-colour diagrams for the early-type galaxies have shown that most of the early type galaxies exhibit an ongoing star formation process with a SFR (star formation rate) in the range $\sim 0.1 - 1 $\msol/yr. This SFR is reproducible from the available matter accumulated through the mass-loss from the evolved stars in these galaxies. However, there must be an additional contribution of the ISM from the merger and/or gas in fall processes in such galaxies.

We estimate the SFR in this galaxy using IRAS flux densities and the relation given by Thronson \& Telesco (1986), and is found to take place at the rate of 0.23\msol\,$yr^{-1}$. Here we would like to mention that this estimate of SFR is inferred from the FIR flux densities and provides an upper limit. 

\section{Results \& Discussion}
\subsection{Results} 
\begin{itemize}
\item Our analysis confirms the presence of a prominent dust lane in NGC 4370 aligned along its optical major axis at PA\,=\,85$^o$ and extended up to 1\,\arcmin. 
\item The extinction curve derived for NGC 4370 is found to run parallel to the Galactic extinction curve, which imply that the properties of dust grains in NGC 4370 are similar to those of canonical grains in the Milky Way. The ratio of the total V band extinction to the selective extinction in B and V bands (the $R_V$ value) is found to be equal to 2.85$\pm$0.05 which is smaller compared to the canonical value of 3.1 in the Milky Way and is consistent with the values reported for dust-lane galaxies by Goudfrooij \etal (1994b), Patil \etal (2007) and also by Finkelman \etal (2008).
\item Dust masses quantified from the total optical extinction measurement and IRAS flux densities are $4.4\times 10^4\, M_\odot$ and $2.0\times 10^5\, M_\odot$, respectively, showing a discrepancy in the two estimates. This discrepancy is due to the fact that the dust mass estimation using optical extinction value critically depends on the spatial distribution of the dust with respect to the stars.
\item Morphology of the "warm gas" derived from the analysis of narrow band image centred on $H_\alpha$ emission matches closely with the dust morphology and suggest a strong association among the two components of ISM.
\item Analysis of 41\,ks {\it Chandra} X-ray image do not show obvious X-ray emission, perhaps due to the poor response of the detector at about $10\arcmin$ off the centre of the ACIS-3 CCD where NGC 4370 was lying. 
\item The SFR derived from the IRAS flux densities is found to be equal to $0.23\, M_\odot / yr$.
\end{itemize}

\begin{figure}[!htb]
\begin{centering}
\includegraphics[width=2.80in]{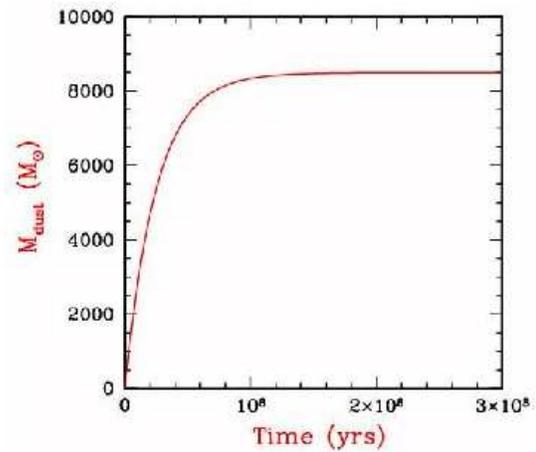}
\caption{Expected total amount of dust that will be accumulated by the galaxy NGC 4370 over its lifetime}
\label{dstbld}
\end{centering}
\end{figure}

\subsection{Discussion}
The issue of origin of dust in this class of galaxies is highly controversial. Previously it was thought that low mass stars, which take around a billion years to evolve, were the main source of the dust budget contributing 90\% of the observed dust in the ISM. However, the timescales involved are far long to explain the observed quantum of dust in these galaxies (Morgan \& Edmunds 2003). For examining internal origin of the dust in NGC 4370, we assume a simple steady state model in which dust is continuously produced by normal mass loss from the evolved stars and is continuously destroyed by ion sputtering. The grain production rate is proportional to the local stellar density and the specific stellar mass-loss rate from the older stellar population $\alpha_\star = \frac{\dot\mathcal{M}_\star}{\mathcal{M}_\star} = 4.7\times 10^{-20}$ s$^{-1}$ (Mathews 1989). The sputtering rate of the grains is determined by the local density and temperature in the hot gas as determined from X-ray flux (Goudfrooij \& de Jong 1995). Using this rate of dust production and simultaneous destruction by sputtering, we compute the total dust accumulated by NGC 4370 over its life time. The total build up of dust mass in NGC 4370 is shown in Figure \ref{dstbld} and is found to be equal to $8.9\times 10^3$ \msol\,  and is much smaller than the dust estimated by optical extinction method. This suggest that at least some fraction of dust is acquired by NGC 4370 externally through a merger like event. Bertola \etal (1988) on the basis of kinematical study have shown that the angular momenta of gas and stars in NGC 4370 are co-rotating and are probably resulted due to the acquisition of external material by this galaxy later in its history. There are similar evidences reported by several other authors. This strongly support that the dust and gas in NGC 4370 have an external origin through a merger like event.\\

{\it Acknowledgment}:
We are thankful of the time-allocation committee of Himalayan Chandra Telescope and the scientific staff of its control station at Hoskotte for making this data available. MKP is also thankful of Honble  Vice-Chancellor of SRTM Univesity for his generous support for the research activity.  MKP is thankful of IUCAA for the usage of the excellent computer and library facilities available at IUCAA. We acknowledge  the use of data from the archives of the Chandra, 2MASS and GALEX.

\end{document}